# Dependence of Mobility on Density of Gap States in Organics by GAMEaS - Gate Modulated Activation Energy Spectroscopy

Woo-young So[1], David V. Lang[1], Vladimir Y. Butko[3,†], Xiaoliu Chi[1,#],

Jason C. Lashley[3], and Arthur P. Ramirez[1, 2]

[1]*Columbia University, New York, NY 10027*
[2]*Bell Laboratories, Lucent Technologies, 600 Mountain Avenue, Murray Hill, NJ 07974*
[3]*Los Alamos National Laboratory, Los Alamos, NM 87545*

ABSTRACT

We develop a broadly applicable transport-based technique, GAte Modulated activation Energy Spectroscopy (GAMEaS), for determining the density of states (DOS) in the energy gap. GAMEaS is applied to field effect transistors made from different single crystal oligomer semiconductors to extract the free-carrier mobility, $\mu_0$, from the field effect mobility, $\mu_{eff}$. Samples with a lower DOS exhibit higher $\mu_{eff}$. Values of $\mu_0$ up to $100 \pm 40$ $cm^2/Vs$ at 300K are observed, showing that performance can be greatly enhanced by improving sample purity and crystal quality.

Recently, organic semiconductors have attracted great interest due to their potential for cost-effective, high-performance devices such as field-effect transistors (FETs), light-emitting diodes (LEDs), and solar cells. The fundamental charge and energy transport mechanisms underlying these device functionalities can best be understood in the relatively pristine environment of crystalline material, where defect types can be studied in isolation. For this purpose, one wants to establish the causal relationship between the density of gap states (DOS), which are induced chemically and/or mechanically, and device transport and optical properties. For inorganic semiconductors the DOS can be determined by deep level transient spectroscopy performed over a wide temperature range, extending well above room temperature for materials with band gaps as wide as typical organics. Organics, however, are van der Waals bonded with relatively low sublimation temperatures and thus dictate a lower-temperature approach.

In this letter we develop a broadly applicable technique which we call GAte Modulated activation Energy Spectroscopy (GAME$_a$S) for quantitatively establishing the DOS versus Fermi energy, $F$, in organics. In GAME$_a$S, the activation energy, $E_a$, of the source-drain current is measured as a function of gate bias, $V_g$, in an FET structure. The DOS as a function of $F$ is then solved in a recursive way by using $E_a = F$ as the trial function.[1] We use GAME$_a$S to characterize FET devices fabricated from single crystals of pentacene, tetracene, and rubrene, which were found to have field-effect mobilities ranging from $\mu_{eff}$ = 0.01 to greater than 10 $cm^2/Vs$ at 300K.[2][3][4][5][6] This is to be compared to typical FET devices based on thin film polycrystalline material where maximum $\mu_{eff}$ values are of order 1 $cm^2/Vs$. Using GAME$_a$S we find that the free-carrier mobility, $\mu_0$, can range as high as 100 $cm^2/Vs$ in our devices. The difference between $\mu_{eff}$ and $\mu_0$ is attributed to the presence of deep gap states, also identified by GAME$_a$S. These results demonstrate the potential for high performance organic devices through materials quality improvement.

The basic methods used by us for organic crystal growth and for device fabrication and measurement have been described previously.[7][8] The device-dependent parameter is the gate capacitance per unit area (tet10 = 4.29; rub19 = 3.75; rub4 = 5.58; ptc45 = 6.20; ptc63 = 5.58; ptc66 = 4.65; all in nF/cm$^2$). In addition, tetracene sample tet10 and rubrene rub19 are synthesized from commercial-grade material while rubrene rub4 is from a special-order high-purity source. Pentacene sample ptc45 is two-times purified, whereas ptc63 and ptc66 have been purified by re-crystallizing four times. We had previously shown in pentacene and tetracene FETs that contact potentials do not dominate the current flow.[7][8] In rubrene FETs contact potentials have been reported to be negligible at high source-drain voltage, $V_{ds}$,[2] and over the temperature range 200$K$ to 300$K$.[9] Thus, we neglect the effect of contact potentials in this work, although they might be important for other studies.

The basic idea of GAME$_a$S is to extract the DOS from the rate of change of $E_a$ versus $V_g$. We, as well as others, have used a simplified version of this analysis, valid in the limit of low free-carrier density by assuming $F = E_a$.[1][10] The simplified analysis breaks down, however, when the free carrier density approaches the total density of induced carriers, as occurs for high $V_g$ and high $\mu_0$. In this regime, $E_a$ becomes a non-trivial function of $F$ requiring a recursive solution, which we describe below.

Quite generally, we know that the total surface charge induced by the gate bias is $n_{total} = n_{free} + n_{trapped}$, the densities of free and trapped charges respectively, and the corresponding gate-induced charge equation can be rewritten in terms of $V_g$ and $F$ as:

$$-CV_g/q\alpha = \int_{E_{v,bottom}}^{E_v} N_{band}(E)(1-f(E,F))dE + \int_{E_v}^{E_c} N_{gap}(E)(1-f(E,F))dE, \quad (1)$$

where $C$ is the gate capacitance per unit area, $\alpha$ is a correction factor described below, $E_{v,bottom}$ is the energy at the bottom of the valence band, $N_{band}(E)$ and $N_{gap}(E)$ are the DOSs per unit area in the valence band and in the gap respectively, and $f$ is the Fermi-Dirac function. In saturation, namely $|V_{ds}| \geq |V_g|$, $\alpha = 2$, reflecting the so-called gradual channel approximation (GCA), i.e. a surface potential varying linearly from the source to the pinch-off position adjacent to the drain.[11] The pinch-off region length in the devices studied here is negligible since the differential impedance in saturation is large, >100MΩ, as expected for long-channel devices. The first term on the right side of Eq. 1 simplifies to $N_o(T/300)exp(-F/kT)$ for $(F-E)>3kT$, where $N_o$ is the two-dimensional effective DOS in the valence band at 300K, estimated to be $1.08 \times 10^{13}\ cm^{-2}$, assuming $m^*/m_o = 1$. The lower limit of the integral in the second term of Eq. 1 can be substituted with $F$ and we assume the zero-temperature approximation of $1-f$, valid when $N_{gap}(E)$ has a slope parameter less than $1/kT$. Thus, differentiating both sides of Eq. 1. we find,

$$N_{gap}(F) = -\frac{d}{dF}\int_F^{E_c} N_{gap}(E)dE \cong (C/q\alpha)(dV_g/dF) - N_o(T/300)exp(-F/kT)/kT. \quad (2)$$

This is the operative expression for extracting $N_{gap}(F)$ from $E_a(V_g)$ data and describes the DOS in most crystalline FET devices.

In order to solve Eq. (2), we need to express $F$ in terms of the measured $E_a$. We start with the Arrhenius form, $E_a/k = -d(lnI_{ds}(T))/d(1/T)$, where the source-drain current, $I_{ds}$, is given by $I_{ds}(T)=A\mu_{o300}(T/300)^m N_o(T/300)exp(-F/kT)$ where $A$ and $m$ are constants, the latter ranging between 2.2 and 2.9 for phonon scattering.[5,12,13] Thus, $E_a$ is given by $E_a = F - (m-1)kT + (1/T)dF/d(1/T)$, or

$$F=E_a+(m-1+\gamma)kT, \qquad (3)$$

where $\gamma(F;N_{gap}) \equiv dF/d(1/kT)$, is a functional of $N_{gap}(E)$.[14] Thus Eq. 3 must be solved numerically using a best estimate of $N_{gap}(E)$.

Implementation of GAME$_a$S to derive the DOS from $E_a(V_g)$ data amounts to solving Eqs. (2) and (3) recursively. As a starting point, $F = E_a$ is used as the trial function, and $N_{gap}(F)$ is derived by Eq (2). For the next determination of $F$, while the exact $N_{gap}(F)$ can be used in Eq. 3, the calculation is greatly simplified by using an analytical form fit to $N_{gap}(F)$. We find little loss of accuracy in approximating $N_{gap}(F)$ by an exponential fit to the low energy tail. The procedure typically converges with 3-5 iterations, yielding a self-consistent DOS. Fig. 1 shows $E_a$ versus $V_g$ for several samples while the inset in Fig. 1 shows the numerically-solved $F(E_a)$ for two different analytic approximations to $N_{gap}(F)$.

The DOS in Fig. 2 is derived from the data in Fig. 1 using the GAME$_a$S method described above. The DOS of our samples at low energy could be approximated by one of two types of exponential band tails – a shallow band tail $N_{gap1}(E) = N_{10}\exp(-\beta_1 E)$ with $\beta_1 = 29eV^{-1}$ and $N_{10}= 7.0 \times 10^{14}\ cm^{-2}eV^{-1}$ or a deep band tail $N_{gap2}(E) = N_{20}\exp(-\beta_2 E)$ with $\beta_2 = 13eV^{-1}$ and $N_{20}= 1.5 \times 10^{14}\ cm^{-2}eV^{-1}$. We note that the slope $\beta_1 = 29eV^{-1}$ in our purest crystals is close to $1/kT$ suggesting that these states are affected by thermal vibrations in the two dimensional FET channel. In general, we see a large variation in the densities of the deep and shallow states among different samples with smaller $N_{gap}$ found in samples that underwent multiple purifications. Sample ptc45 has a peak at $0.43eV$, which may be due to a hydrogen-induced defect,[15] and tet10 has a deep band tail similar to ptc45 without the peak at $0.43eV$. Sample rub19 also falls on the same deep band tail as ptc45 and tet10. In contrast, ptc63, ptc66, and

rub4, which are highly pure, have nearly an order of magnitude less of the deep band tail $N_{gap2}$. These results bear some similarity to observations in low-$\mu_{eff}$ polycrystalline TFTs, where Völkel et al[16] used a device model at 300K to obtain shallow and deep exponential band tails with $\mu_{eff}$ and deep band tail DOS being sample dependent. The present results show, however, that it is possible to eliminate the deep band tail, $N_{gap2}$, in single crystals by using iterative purification methods. Importantly, $\mu_{eff}$ also depends on sample purification. Sample ptc45 is purified only two times and has $\mu_{eff} = 0.3\ cm^2/Vs$,[7] whereas four-time crystallized ptc66 and ptc63 have $\mu_{eff} = 2.0$ and $2.5\ cm^2/Vs$, respectively. Commercial-grade tet10 and rub19 have $\mu_{eff} = 0.15$ and $4.3\ cm^2/Vs$ respectively, while highly-purified rub4 has $\mu_{eff} = 12\ cm^2/Vs$. As expected, $\mu_{eff}$ increases for higher purity samples. The GAME$_a$S analysis presented here shows that the increase in $\mu_{eff}$ correlates with a measured decrease in the DOS.

It is interesting to compare the surface DOS in FETs to the bulk DOS that we found by measuring the photoquenching rate previously for pentacene crystals.[1] In order to evaluate a surface density of the bulk DOS, we introduce two effective depths: one is the monolayer scale, $1nm$,[17] and the other is an extended depth of 10nm. Thus, the converted $N_{b,1nm}\ (E)$ or $N_{b,10nm}\ (E)$ can be expressed by $N_{b0} \exp(-\beta_{bulk} E) \times d$, where $d$ is $1nm$ or $10\ nm$ respectively with $\beta_{bulk}= 5.7 eV^{-1}$ and $N_{b0}= 5.5 \times 10^{18}\ cm^{-3} eV^{-1}$, and they are also plotted in the Fig. 2. The DOSs observed here are much higher than the reported bulk value suggesting that the FET band tails reflect the characteristics of the dielectric/crystal interface, and not the bulk single crystal.

We now show how to use the DOS and $F$ determined using GAME$_a$S to extract $\mu_o$ from $\mu_{eff}$. As suggested in mobility edge model,[10] $\mu_{eff} = \mu_o(T/300)^{-m}\Theta(F,T)$, where $\Theta \equiv n_{free}/n_{total}$ and $n_{free}$ and $n_{total}$ are given by $n_{free} = N_o(T/300)\exp(-F/kT)$ and $n_{total} = CV_g/q\alpha$, where $\alpha$ is unity for the linear condition, $|V_{ds}| << |V_g|$. In Figure 3 is shown $\mu_{eff}$, as derived from transfer curves, versus $\mu_o$ calculated through the above relationship; $F$ is corrected as above from the measured $E_a(V_g=-45, -50V)$ and $\alpha=2$ since our six devices are operated at saturation. We also include in Fig. 3 temperature dependent FET data from the literature,[2 3 4 9 18 19 20] with the same assumptions for $\gamma$ and $m$ as above. In order to evaluate $\Theta$ for the literature data, we use the reported capacitance, a corrected $F$, $\alpha$ as appropriate, and the $V_g$ at which $E_a$ is measured. Note that since we do not know the device-dependent parameters or $E_a(V_g)$ for the literature data we must assume a DOS, and for this purpose, use $N_{gap1}$ and $N_{gap2}$ as above. The shaded region in Fig. 3 corresponds to the physically inaccessible condition $\Theta > 1$.

The data in Fig. 3 show that, whereas $\mu_{eff}$ data taken from a variety of samples and types of FET architectures has a wide variation, the values of $\mu_0$ extracted using GAME$_a$S collapse into a narrow range of values substantially higher than presently measured $\mu_{eff}$ values. In addition, the $\mu_o$ of thin-film devices is found to be lower than in crystalline devices by about one order of magnitude, demonstrating that $\mu_o$ depends on the crystallinity of the underlying materials. The narrowness of the range of derived $\mu_o$ values suggests that this quantity is not strongly dependent on defects introduced by the FET fabrication method. For example, the parylene-gate FETs[9] of the Rutgers group have a relatively low $\mu_{eff} = 0.20$ $cm^2/Vs$ whereas their elastomer stamp FETs[3] have 7.5 $cm^2/Vs$. Nevertheless the two devices have similar free-carrier mobilities: 11 and 21 $cm^2/Vs$ respectively.

The relationships found here among $\mu_o$, the DOS, crystal purification, and processing conditions shows quantitatively that the $\mu_{eff}$ of present-day single-crystal FET structures is not intrinsic and can be further improved by advances in crystallization and purity; in other words, higher $\mu_{eff}$ can be achieved by increasing $\Theta$ with fewer trapping states as indicated in Fig. 3. This is corroborated by the fact that the DOS in the high-mobility pentacene devices obtained here is still significantly higher than in bulk pentacene crystals.[1] This relationship also explains the apparent discrepancy between the bulk (space-charge limited current) mobility measurement of $\mu_o = 35$ $cm^2/Vs$ at 300K with no apparent carrier trapping in highly purified pentacene crystals[5] and the much lower effective mobility (2.0-2.5 $cm^2/Vs$) of highly purified pentacene FETs reported here. As is clear from Fig. 3, both of these systems have nearly the same free-carrier mobility $\mu_o$ but the FETs have a lower $\Theta$ due to more localized states near the dielectric interface. Thus the preponderance of data suggests that the intrinsic hole mobility at 300K in single crystal polyacenes can be pushed up to 100 cm$^2$/Vs. The challenge for both crystal growth and device fabrication is, therefore, to minimize impurities and process-induced gap states so that such mobility values can be realized in the performance of actual FETs.

In summary, we present an easily-realized method to extract the approximate DOS from $I_{ds}(T)$ versus $V_g$ and show that the limiting exponential band tail in purified rubrene and pentacene crystal-based FETs has a slope parameter of $1/kT$ at 300K. The field-effect mobility is fully parameterized by two factors: (1) the mobility of free carriers in the FET channel and (2) the degree of carrier trapping in localized band-tail states, which are much higher in concentration for FET structures than for bulk crystals. We show that state-of-the-art crystalline FETs fabricated from rubrene, pentacene, and tetracene all demonstrate much higher intrinsic mobilities than previously measured via FET characteristics and that further improvement in FET mobilities will be possible with higher quality single crystal. Such improvements will involve removal of impurities such as dihydropentacene and pentacenequinone from pentacene, the removal of rubrene derivatives from rubrene, and the reduction of surface trap state densities through FET process improvement. The GAME$_a$S technique might provide useful feedback for such studies.

We acknowledge the support of the US Department of Energy under grant DE-FG02-04ER46118 and the Laboratory Directed Research and Development Program at Los Alamos National Laboratory. This work was partially supported by the Nanoscale Science and Engineering Initiative of the National Science Foundation under NSF Award Number CHE-0117752 and by the New York State Office of Science, Technology, and Academic Research (NYSTAR).

[†] Present address, Brookhaven National Laboratory.
# Present address, Texas A&M University, Kingsville, TX.

Figure Captions

Figure 1. Measured $E_a$ in the temperature range 200-300$K$ versus $V_g$ for three pentacene crystals (ptc63, ptc66, and ptc45[7]), two rubrene crystals (rub4 and rub19), and one tetracene crystal (tet10[8]). The inset shows the correlation between measured activation energy and corrected Fermi energy at 300$K$ for DOS which are assumed in this letter.

Figure 2. Densities of localized states in the gap obtained from $E_a$ versus $V_g$ for the six crystalline FETs in Fig. 1. The energy is measured from the top of the valence band and is corrected to approximate the Fermi energy $F$ to within +/-0.01$eV$. The crystalline pentacene bulk density of states is from ref. 7.

Figure 3. Effective FET hole mobility at room temperature versus the intrinsic mobility derived using GAME$_a$S. Current work: ptc45, ptc63, ptc66, rub4, rub19, tet10; literature: ptc-P1,2,[19] ptc-E1,[20] rub-R1,[9] rub-R2,[3] tet-D1,2[18].

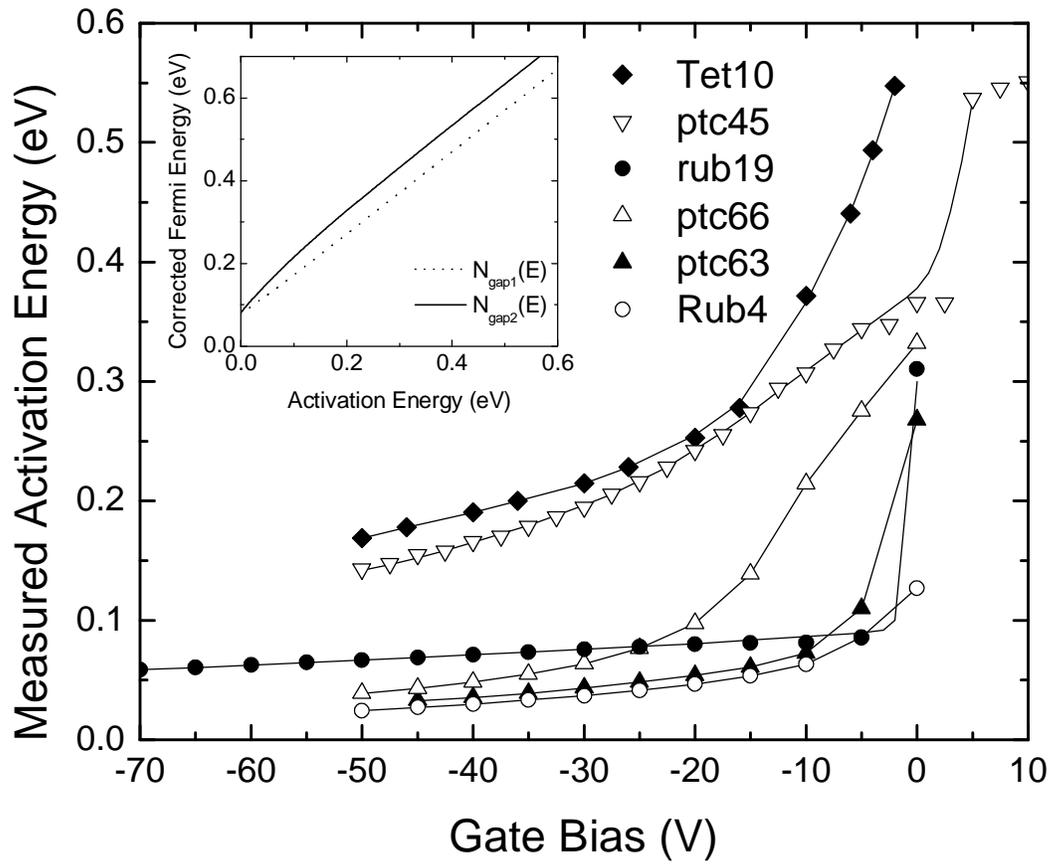

**Fig. 1**

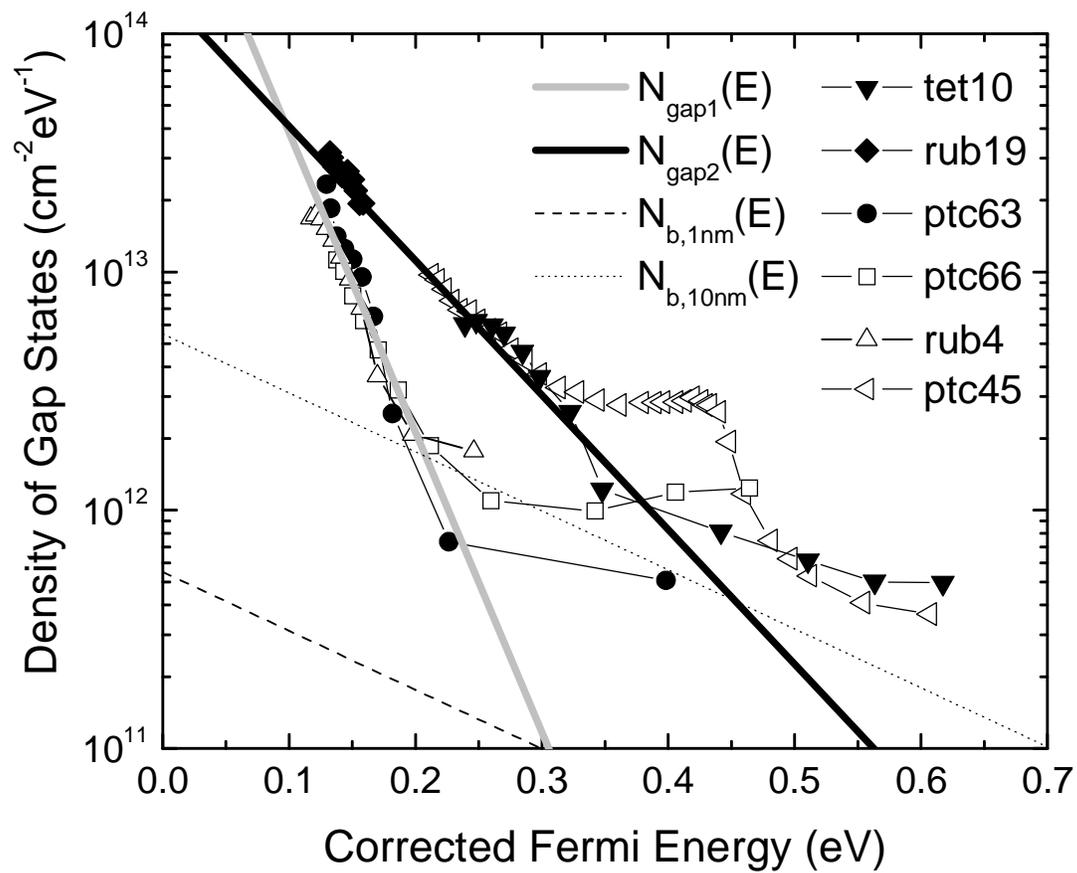

**Fig. 2**

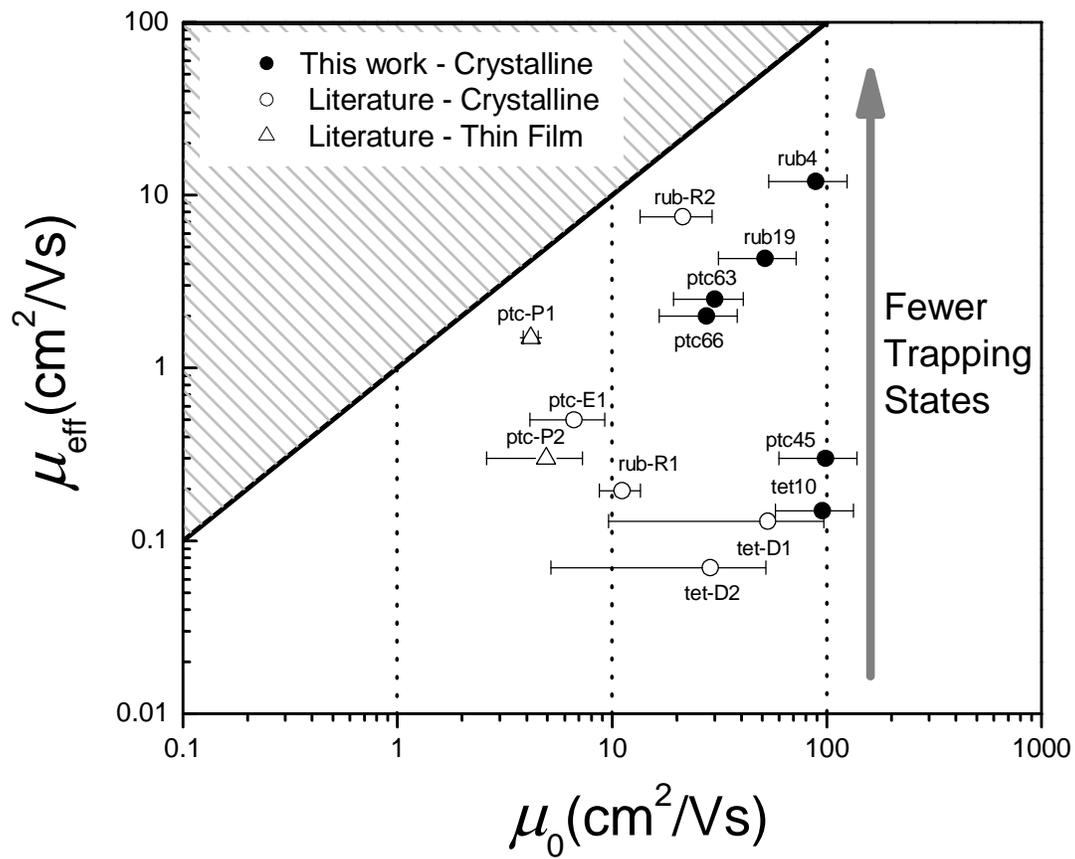

**Fig. 3**